# Generalized Maiorana-McFarland Constructions for Almost Optimal Resilient Functions

WeiGuo Zhang and GuoZhen Xiao
(ISN Lab, Xidian University, Xi'an 710071, China)

**Abstract**

In a recent paper [1], Zhang and Xiao describe a technique on constructing almost optimal resilient functions on even number of variables. In this paper, we will present an extensive study of the constructions of almost optimal resilient functions by using the generalized Maiorana-McFarland (GMM) construction technique. It is shown that for any given $m$, it is possible to construct infinitely many $n$-variable ($n$ even), $m$-resilient Boolean functions with nonlinearity equal to $2^{n-1} - 2^{n/2-1} - 2^{k-1}$ where $k < n/2$. A generalized version of GMM construction is further described to obtain almost optimal resilient functions with higher nonlinearity. We then modify the GMM construction slightly to make the constructed functions satisfying strict avalanche criterion (SAC). Furthermore we can obtain infinitely many new resilient functions with nonlinearity $> 2^{n-2} - 2^{(n-1)/2}$ ($n$ odd) by using Patterson-Wiedemann functions or Kavut-Yücel functions. Finally, we provide a GMM construction technique for multiple-output almost optimal $m$-resilient functions $F: \mathbb{F}_2^n \mapsto \mathbb{F}_2^r$ ($n$ even) with nonlinearity $> 2^{n-1} - 2^{n/2}$. Using the methods proposed in this paper, a large class of previously unknown cryptographic resilient functions are obtained.

## 1 Introduction

Confusion and diffusion, introduced by Shannon [2], are two important principles used in the design of symmetric cryptosystems (stream ciphers and block ciphers). Boolean functions possessing multiple cryptographic criteria play an important role in enforcing these principles. The following criteria for cryptographic Boolean functions are often considered: high nonlinearity, high resiliency, high algebraic degree and strict avalanche criterion (SAC). The tradeoffs among these criteria are difficult problems and have received lots of attention. By an $(n, m, d, N_f)$ function we mean an $n$-variable, $m$-resilient Boolean function $f$ with algebraic degree $d$ and nonlinearity $N_f$. Siegenthaler [3] and Xiao [4] proved that $d \leq n - m - 1$ for $n$-variable, $m$-resilient functions. Such a function, reaching this bound, is called degree-optimized. For relations between SAC and resiliency, one can find in [5], [6].

Construction of resilient functions with high nonlinearity has been a challenging research problem in cryptography for twenty years[7][8][9][10][11][12][13][14][1]. On even number of variables $n$, Bent functions [15] achieve optimal nonlinearity $2^{n-1} - 2^{n/2-1}$, but they are not resilient and their algebraic degrees are not more than $n/2$. For the case when $n \geq 9$ is odd, the maximum achievable value of $N_f$ is unknown in general, and we know only that it is strictly larger than $2^{n-1} - 2^{(n-1)/2}$ [16]. (For odd $n \leq 7$, the optimal nonlinearity of $n$-variable functions is $2^{n-1} - 2^{(n-1)/2}$.) An $n$-variable Boolean function $f$ is said to be almost optimal if $N_f \geq 2^{n-1} - 2^{\lfloor n/2 \rfloor}$. The problem how tight is the nonlinearity bound of resilient Boolean functions remains open. Construction of almost optimal resilient functions has been discussed in [11], [12], [13], [14], [1], and will also be extensive studied in this paper.

A classical class of cryptographic Boolean functions are the Maiorana-McFarland (M-M) class which can ensure many of the criteria above mentioned. For more detailed information about M-M class functions please see [17][1] and their references. In this paper, we will introduce a generalized Maiorana-McFarland (GMM) construction technique to obtain almost optimal resilient functions.

The organization of this paper is as follows. In Section 2, the basic concepts and notions are presented. Section 3 describes the GMM construction technique. The resilient functions satisfying SAC with very high nonlinearity are constructed. The degree of the GMM type resilient functions can also be optimized. In Section 4, by using Patterson-Wiedemann functions or Kavut-Yücel functions, many new $n$-variable resilient functions with nonlinearity $> 2^{n-2} - 2^{(n-1)/2}$ ($n$ odd) are obtained. In section 5, we provide a construction technique for multiple-output resilient functions on $n$ variables ($n$ even) with nonlinearity $> 2^{n-1} - 2^{n/2}$. Section 6 concludes the paper with several open problems.

## 2 Preliminary

Let $\mathcal{B}_n$ denote the set of Boolean functions of $n$ variables. A Boolean function $f(X_n) \in \mathcal{B}_n$ is a function from $\mathbb{F}_2^n$ to $\mathbb{F}_2$, where $X_n = (x_1, \cdots, x_n) \in \mathbb{F}_2^n$ and $\mathbb{F}_2^n$ is the vector space of tuples of elements from $\mathbb{F}_2$. To avoid confusion with the additions of integers in $\mathbb{R}$, denoted by $+$ and $\Sigma_i$, we denote the additions over $\mathbb{F}_2$ by $\oplus$ and $\bigoplus_i$. For simplicity, we denote by $+$ the addition of vectors of $\mathbb{F}_2^n$. $f(X_n)$ is generally represented by its algebraic normal form (ANF):

$$f(X_n) = \bigoplus_{u \in \mathbb{F}_2^n} \lambda_u (\prod_{i=1}^{n} x_i^{u_i}) \tag{1}$$

where $\lambda_u \in \mathbb{F}_2$, $u = (u_1, \cdots, u_n)$. The algebraic degree of $f(X_n)$, denoted by $deg(f)$, is the maximal value of $wt(u)$ such that $\lambda_u \neq 0$, where $wt(u)$ denotes the Hamming weight of $u$. $f$ is called an affine function when $deg(f) = 1$. An affine function with constant term equal to zero is called a linear function. Any linear function on $\mathbb{F}_2^n$ is denoted by:

$$\omega \cdot X_n = \omega_1 x_1 \oplus \cdots \oplus \omega_n x_n,$$

where $\omega = (\omega_1, \cdots, \omega_n)$, $X_n = (x_1, \cdots, x_n) \in \mathbb{F}_2^n$. The Walsh spectrum of $f \in \mathcal{B}_n$ in point $\omega$ is denoted by $W_f(\omega)$ and calculated by

$$W_f(\omega) = \sum_{X_n \in \mathbb{F}_2^n} (-1)^{f(X_n) \oplus \omega \cdot X_n}. \tag{2}$$

$f \in \mathcal{B}_n$ is said to be balanced if its output column in the truth table contains equal number of 0's and 1's (i.e. $W_f(0) = 0$).

In [4], a spectral characterization of resilient functions has been presented.

***Lemma 1:*** A $n$-variable Boolean function is $m$-resilient if and only if its Walsh transform satisfies

$$W_f(\omega) = 0, \text{ for } 0 \leq wt(\omega) \leq m, \omega \in \mathbb{F}_2^n. \tag{3}$$

In term of Walsh spectra, the nonlinearity of $f \in \mathcal{B}_n$ is given by [18]

$$N_f = 2^{n-1} - \frac{1}{2} \cdot \max_{\omega \in \mathbb{F}_2^n} |W_f(\omega)|. \tag{4}$$

***Definition 1:*** The Boolean function $f \in \mathcal{B}_n$ is said to be almost optimal if

- $N_f \geq 2^{n-1} - 2^{n/2}$, when $n$ is even;
- $N_f \geq 2^{n-1} - 2^{(n-1)/2}$, when $n$ is odd.

The autocorrelation function of $f \in \mathcal{B}_n$ is defined by

$$\Delta_f(\alpha) = \sum_{X_n \in \mathbb{F}_2^n} (-1)^{f(X_n) \oplus f(X_n + \alpha)} \tag{5}$$

The SAC was introduced by Webster and Tavares [19]. $f$ satisfies SAC if

$$\Delta_f(\alpha) = 0, \quad \text{for } wt(\alpha) = 1. \tag{6}$$

***Definition 2 ([7]):*** The functions of original M-M class are defined as follows: For any positive integers $p$, $q$ such that $n = p + q$ an M-M function is a function $f \in \mathcal{B}_n$ defined by

$$f(Y_q, X_p) = \phi(Y_q) \cdot X_p \oplus \pi(Y_q), \qquad X_p \in \mathbb{F}_2^p, Y_q \in \mathbb{F}_2^q \tag{7}$$

where $\phi$ is any mapping from $\mathbb{F}_2^q$ to $\mathbb{F}_2^p$ and $\pi \in \mathcal{B}_q$.

## 3 GMM construction

This section presents two versions of GMM construction methods for constructing almost optimal resilient functions. The SAC and degree optimization of the GMM type functions are also considered.

### 3.1 A reduced version

***Construction 1:*** Let $n \geq 12$ be even, and let $m$ be a positive integer such that there exists an integer $k$ with

$$k = \min_{m < s < n/2} \{s \mid 2^{n/2 - s} \cdot \sum_{i=0}^{m} \binom{n/2}{i} \leq \sum_{j=m+1}^{s} \binom{s}{j}\}. \tag{8}$$

Let

$$T_0 = \{a \mid wt(a) > m, \ a \in \mathbb{F}_2^{n/2}\} \tag{9}$$

and

$$T_1 = \{c \mid wt(c) > m, \ c \in \mathbb{F}_2^k\}. \tag{10}$$

Let $E_0$ be any subset of $\mathbb{F}_2^{n/2}$ with

$$|E_0| = \sum_{i=m+1}^{n/2} \binom{n/2}{i} = |T_0| \tag{11}$$

Let $\overline{E_0} = \mathbb{F}_2^{n/2} \setminus E_0$ and $E_1 = \overline{E_0} \times \mathbb{F}_2^{n/2-k}$. Denote by $\phi_0$ any bijective mapping from $E_0$ to $T_0$, $\phi_1$ any injective mapping from $E_1$ to $T_1$. Let $X_n = (x_1, \cdots, x_n) \in \mathbb{F}_2^n$, $X_t' = (x_1, \cdots, x_t) \in \mathbb{F}_2^t$ and $X_{n-t}'' = (x_{t+1}, \cdots, x_n) \in \mathbb{F}_2^{n-t}$ where $t \in \{n/2, k\}$. Then we construct the function $f \in \mathcal{B}_n$ as follows:

$$f(X_n) = \begin{cases} \phi_0(X_{n/2}') \cdot X_{n/2}'' & \text{if } X_{n/2}' \in E_0 \\ \phi_1(X_{n-k}') \cdot X_k'' & \text{if } X_{n-k}' \in E_1. \end{cases} \tag{12}$$

**Remark:** For Inequality (8) holds, we always have $|E_1| \leq |T_1|$. So we can find an injective mapping $\phi_1$.

**Theorem 1:** Let $f \in \mathbb{F}_2^n$ be as in Construction 1. Then $f$ is an almost optimal $(n, m, d, N_f)$ function with

$$N_f = 2^{n-1} - 2^{n/2-1} - 2^{k-1}. \tag{13}$$

Let $\phi_1(X_{n-k}')_j$ be the $j$-th component of $\phi_1(X_{n-k}')$, where $1 \leq j \leq k$. If

$$\bigoplus_{X_{n-k}' \in E_1} \phi_1(X_{n-k}')_j = 1 \tag{14}$$

for some $j$, then $d = n - k + 1$.

**Proof:** For any $\omega = (\omega_1, \cdots, \omega_n) \in \mathbb{F}_2^n$ we have

$$W_f(\omega) = \sum_{X_n \in \mathbb{F}_2^n} (-1)^{f(X_n) \oplus \omega \cdot X_n} = S_0 + S_1 \tag{15}$$

where

$$S_0 = \sum_{X_{n/2}' \in E_0} (-1)^{(\omega_1, \cdots, \omega_{n/2}) \cdot X_{n/2}'} \sum_{X_{n/2}'' \in \mathbb{F}_2^{n/2}} (-1)^{(\phi_0(X_{n/2}') + (\omega_{n/2+1}, \cdots, \omega_n)) \cdot X_{n/2}''} \tag{16}$$

$$S_1 = \sum_{X_{n-k}' \in E_1} (-1)^{(\omega_1, \cdots, \omega_{n-k}) \cdot X_{n-k}'} \sum_{X_k'' \in \mathbb{F}_2^k} (-1)^{(\phi_1(X_{n-k}') + (\omega_{n-k+1}, \cdots, \omega_n)) \cdot X_k''} \tag{17}$$

Case 1: $0 \leq wt\left((\omega_{n/2+1}, \cdots, \omega_n)\right) \leq m$.

Since $\phi_0$ is a mapping from $E_0$ to $T_0$, from (9), we have $wt(\phi_0(X_{n/2}')) \geq m+1$. Obviously, $\phi_0(X_{n/2}') + (\omega_{n/2+1}, \cdots, \omega_n) \neq \mathbf{0}$. Thus

$$\sum_{X_{n/2}'' \in \mathbb{F}_2^{n/2}} (-1)^{(\phi_0(X_{n/2}') + (\omega_{n/2+1}, \cdots, \omega_n)) \cdot X_{n/2}''} = 0 \tag{18}$$

We obtain $S_0 = 0$. Similarly, for $0 \leq wt\left((\omega_{n-k+1}, \cdots, \omega_n)\right) \leq m$ and $wt(\phi_1(X_{n-k}')) \geq m+1$, we have $\phi_1(X_{n-k}') + (\omega_{n-k+1}, \cdots, \omega_n) \neq \mathbf{0}$. we have

$$\sum_{X_k'' \in \mathbb{F}_2^k} (-1)^{(\phi_1(X_{n-k}') + (\omega_{n-k+1}, \cdots, \omega_n)) \cdot X_k''} = 0 \tag{19}$$

Thus, $S_1 = 0$. Then we have $W_f(\omega) = 0$.

Obviously, when $0 \leq wt(\omega) \leq m$, we always have $0 \leq wt(\omega_{n/2+1}, \cdots, \omega_n) \leq m$. Hence, $W_f(\omega) = 0$. By Lemma 1, $f$ is an $m$-resilient function.

Case 2: $wt\left((\omega_{n/2+1}, \cdots, \omega_n)\right) \geq m + 1$.

In this case, for $\phi_0$ is a bijective mapping from $E_0$ to $T_0$, we have $\phi_0^{-1}(\omega_{n/2+1}, \cdots, \omega_n) \in E_0$. When $X'_{n/2} = \phi_0^{-1}(\omega_{n/2+1}, \cdots, \omega_n)$, we have

$$\phi_0(X'_{n/2}) + (\omega_{n/2+1}, \cdots, \omega_n) = 0. \tag{20}$$

Then

$$\sum_{X''_{n/2} \in \mathbb{F}_2^{n/2}} (-1)^{(\phi_0(X'_{n/2}) + (\omega_{n/2+1}, \cdots, \omega_n)) \cdot X''_{n/2}} = 2^{n/2} \tag{21}$$

When $X'_{n/2} \neq \phi_0^{-1}(\omega_{n/2+1}, \cdots, \omega_n)$, we have

$$\sum_{X''_{n/2} \in \mathbb{F}_2^{n/2}} (-1)^{(\phi_0(X'_{n/2}) + (\omega_{n/2+1}, \cdots, \omega_n)) \cdot X''_{n/2}} = 0. \tag{22}$$

Hence,

$$S_0 = (-1)^{(\omega_1, \cdots, \omega_{n/2}) \cdot \phi_0^{-1}(\omega_{n/2+1}, \cdots, \omega_n)} \cdot 2^{n/2} \in \{\pm 2^{n/2}\} \tag{23}$$

Since $\phi_1$ is an injective mapping from $E_1$ to $T_1$, we have

$$\sum_{X''_k \in \mathbb{F}_2^k} (-1)^{(\phi_1(X'_{n-k}) + (\omega_{n-k+1}, \cdots, \omega_n)) \cdot X''_k} = \begin{cases} 0 & \text{if } X'_{n-k} \neq \phi_1^{-1}(\omega_{n-k+1}, \cdots, \omega_n) \\ 2^k & \text{if } X'_{n-k} = \phi_1^{-1}(\omega_{n-k+1}, \cdots, \omega_n). \end{cases} \tag{24}$$

Hence,

$$S_1 \in \{0, \pm 2^k\} \tag{25}$$

Then we have

$$W_f(\omega) \in \{0, \pm 2^{n/2}, \pm(2^{n/2} - 2^k), \pm(2^{n/2} + 2^k)\}. \tag{26}$$

Obviously,

$$\max_{\omega \in \mathbb{F}_2^n} |W_f(\omega)| = 2^{n/2} + 2^k \tag{27}$$

From (4), $f$ is almost optimal with $N_f = 2^{n-1} - 2^{n/2-1} - 2^{k-1}$.

If the equality (14) holds, then the term $x_1 x_2 \cdots x_{n-k} x_{n-k+j}$ will appear in the ANF of $f$. Hence, $deg(f) = n - k + 1$. □

Using the method proposed in Construction 1, for the first time, the almost optimal resilient functions proposed in Table 1 can be constructed. A list of more examples and corresponding cryptographic parameters can be found in Appendix 1 and Appendix 2.

Table 1: Achieved $N_f$ for $n$-variable, $m$-resilient functions

| $n$ | $m$ | $N_f$ | $n$ | $m$ | $N_f$ | $n$ | $m$ | $N_f$ |
|---|---|---|---|---|---|---|---|---|
| 24 | 1 | $2^{23} - 2^{11} - 2^{7}$ | 62 | 2 | $2^{61} - 2^{30} - 2^{19}$ | 42 | 5 | $2^{41} - 2^{20} - 2^{17}$ |
| 28 | 1 | $2^{27} - 2^{13} - 2^{8}$ | 88 | 2 | $2^{87} - 2^{43} - 2^{26}$ | 74 | 5 | $2^{73} - 2^{36} - 2^{27}$ |
| 54 | 1 | $2^{53} - 2^{26} - 2^{15}$ | 20 | 3 | $2^{19} - 2^{9} - 2^{8}$ | 52 | 7 | $2^{51} - 2^{25} - 2^{22}$ |
| 58 | 1 | $2^{58} - 2^{28} - 2^{16}$ | 36 | 3 | $2^{35} - 2^{17} - 2^{13}$ | 70 | 8 | $2^{69} - 2^{34} - 2^{29}$ |
| 30 | 2 | $2^{29} - 2^{14} - 2^{10}$ | 62 | 3 | $2^{61} - 2^{30} - 2^{21}$ | 62 | 9 | $2^{61} - 2^{30} - 2^{27}$ |
| 44 | 2 | $2^{43} - 2^{21} - 2^{14}$ | 92 | 3 | $2^{91} - 2^{45} - 2^{29}$ | 74 | 10 | $2^{73} - 2^{36} - 2^{32}$ |

## 3.2 Generalized version

Let for $1 \leq i \leq n-1$, $E_i \subseteq \mathbb{F}_2^{n-i}$ and $E_i' = E_i \times \mathbb{F}_2^i$ such that

$$\bigcup_{i=1}^{n-1} E_i' = \mathbb{F}_2^n \tag{28}$$

and

$$E_{i_1}' \cap E_{i_2}' = \emptyset, \quad 1 \leq i_1 < i_2 \leq n-1. \tag{29}$$

Let $g_i \in \mathcal{B}_{n-i}$ and $\phi_i$ be a mapping from $\mathbb{F}_2^{n-i}$ to $\mathbb{F}_2^i$. Let

$$X_n = (x_1, \cdots, x_n) \in \mathbb{F}_2^n,$$
$$X_i' = (x_1, \cdots, x_i) \in \mathbb{F}_2^i$$

and

$$X_{n-i}'' = (x_{i+1}, \cdots, x_n) \in \mathbb{F}_2^{n-i}.$$

A cryptographic Boolean function can be constructed as follows:

$$f(X_n) = \phi_i(X_{n-i}') \cdot X_i'' \oplus g_i(X_{n-i}'), \text{ if } X_{n-i}' \in E_i, \ i = 1, 2, \cdots, n-1 \tag{30}$$

where $g_i \in \mathcal{B}_{n-i}$. If for $1 \leq i \leq n-1$, $\phi_i$ is injective mapping and

$$|E_i| \leq \sum_{j=m+1}^{i} \binom{i}{j},$$

then $f$ is an $(n, m, -, N_f)$ function with

$$N_f = 2^{n-1} - \sum_{i=1}^{n-1} a_i 2^{i-1} \tag{31}$$

where

$$a_i = \begin{cases} 0 & \text{if } E_i = \emptyset \\ 1 & \text{if } E_i \neq \emptyset. \end{cases} \tag{32}$$

Especially, when for $n/2 + 1 \leq i \leq n-1$, $E_i = \emptyset$, we always have $N_f > 2^{n-1} - 2^{n/2}$.

Using the generalized version of GMM construction, we can provide functions having parameters which cannot be constructed using the reduced version. Examples for resilient functions which were not known earlier can be found in Table 2.

Table 2: Examples for $(n, m, d, N_f)$ resilient functions which were not known earlier

| | |
|---|---|
| $(32, 1, 26, 2^{31} - 2^{15} - 2^9 - 2^7 - 2^6)$ | $(96, 3, 67, 2^{95} - 2^{47} - 2^{30} - 2^{29})$ |
| $(36, 1, 27, 2^{35} - 2^{17} - 2^{10} - 2^9)$ | $(30, 4, 19, 2^{29} - 2^{14} - 2^{12} - 2^{11})$ |
| $(62, 1, 53, 2^{61} - 2^{30} - 2^{17} - 2^{10} - 2^9)$ | $(36, 4, 24, 2^{35} - 2^{17} - 2^{14} - 2^{12})$ |
| $(66, 1, 55, 2^{65} - 2^{32} - 2^{18} - 2^{16} - 2^{11})$ | $(52, 4, 34, 2^{51} - 2^{25} - 2^{19} - 2^{18})$ |
| $(70, 1, 52, 2^{69} - 2^{34} - 2^{19} - 2^{18})$ | $(62, 4, 41, 2^{61} - 2^{30} - 2^{22} - 2^{21})$ |
| $(74, 1, 55, 2^{73} - 2^{36} - 2^{20} - 2^{19})$ | $(72, 4, 51, 2^{71} - 2^{35} - 2^{25} - 2^{22} - 2^{21})$ |
| $(20, 2, 15, 2^{19} - 2^9 - 2^7 - 2^6)$ | $(86, 4, 59, 2^{85} - 2^{42} - 2^{29} - 2^{27})$ |
| $(34, 2, 24, 2^{33} - 2^{16} - 2^{11} - 2^{10})$ | $(40, 6, 25, 2^{39} - 2^{19} - 2^{17} - 2^{15})$ |
| $(48, 2, 34, 2^{47} - 2^{23} - 2^{15} - 2^{14})$ | $(46, 6, 28, 2^{45} - 2^{22} - 2^{19} - 2^{18})$ |
| $(66, 2, 47, 2^{65} - 2^{32} - 2^{20} - 2^{19})$ | $(52, 6, 32, 2^{51} - 2^{25} - 2^{21} - 2^{20})$ |
| $(70, 2, 50, 2^{69} - 2^{34} - 2^{21} - 2^{20})$ | $(58, 6, 36, 2^{57} - 2^{28} - 2^{23} - 2^{22})$ |
| $(92, 2, 67, 2^{91} - 2^{45} - 2^{27} - 2^{25})$ | $(44, 7, 26, 2^{43} - 2^{21} - 2^{19} - 2^{18})$ |
| $(96, 2, 69, 2^{95} - 2^{47} - 2^{28} - 2^{27})$ | $(58, 7, 38, 2^{57} - 2^{28} - 2^{24} - 2^{22} - 2^{20})$ |
| $(100, 2, 72, 2^{99} - 2^{49} - 2^{29} - 2^{28})$ | $(70, 7, 44, 2^{69} - 2^{34} - 2^{28} - 2^{26})$ |
| $(30, 3, 20, 2^{29} - 2^{14} - 2^{11} - 2^{10})$ | $(56, 8, 33, 2^{55} - 2^{27} - 2^{24} - 2^{23})$ |
| $(46, 3, 33, 2^{45} - 2^{22} - 2^{16} - 2^{13})$ | $(54, 9, 31, 2^{53} - 2^{26} - 2^{24} - 2^{23})$ |
| $(60, 3, 41, 2^{59} - 2^{29} - 2^{20-2^{19}})$ | $(68, 9, 40, 2^{67} - 2^{33} - 2^{29} - 2^{28})$ |
| $(74, 3, 52, 2^{73} - 2^{36} - 2^{24} - 2^{22})$ | $(66, 10, 38, 2^{65} - 2^{32} - 2^{29} - 2^{28})$ |
| $(78, 3, 54, 2^{77} - 2^{38} - 2^{25} - 2^{24})$ | $(86, 10, 51, 2^{85} - 2^{42} - 2^{36} - 2^{35})$ |

## 3.3 SAC

To the best of our knowledge, the nonlinearity values of the known constructed resilient function satisfying SAC are not more than $2^{n-1} - 2^{\lfloor n/2 \rfloor}$ [20][11]. In this section, we present a method to obtain GMM type resilient functions satisfying SAC with nonlinearity $> 2^{n-1} - 2^{\lfloor n/2 \rfloor}$.

**Construction 2:** Let $n \geq 12$ be even, and let $m$ be a positive integer such that there exists an integer $k'$ with

$$k' = \min_{m < s < n/2} \{s \mid 2^{n/2-s+1} \cdot \sum_{i=0}^{m} \binom{n/2}{i} \leq \sum_{j=m+1}^{s-m+1} \binom{s}{j}\}. \tag{33}$$

Let

$$\Gamma_0 = \{a \mid m < wt(a) < n/2 - m, \ a \in \mathbb{F}_2^{n/2}\} \tag{34}$$

and

$$\Gamma_1 = \{c \mid m < wt(c) < n/2 - m, \ c \in \mathbb{F}_2^k\}. \tag{35}$$

Let $\Re_0$ be any subset of $\mathbb{F}_2^{n/2}$ with

$$|\Re_0| = \sum_{i=m+1}^{n/2} \binom{n/2}{i} = |\Gamma_0| \tag{36}$$

Let $\overline{\Re_0} = \mathbb{F}_2^{n/2} \setminus \Re_0$ and $\Re_1 = \overline{\Re_0} \times \mathbb{F}_2^{n/2-k}$. Let $\Omega \subseteq \Gamma_1$ with $|\Omega| = |\Re_1|$ and for any $\beta \in \Omega$, $\beta^c \in \Omega$, where $\beta^c$ is the complementary vector of $\beta$, i.e. $\beta + \beta^c = (11\cdots 1)$. Denote by $\psi_0$ any bijective mapping from $\Re_0$ to $\Gamma_0$, $\psi_1$ any bijective mapping from $\Re_1$ to $\Omega$. Then we construct the function $f' \in \mathcal{B}_n$ as follows:

$$f'(X_n) = \begin{cases} \psi_0(X'_{n/2}) \cdot X''_{n/2} & \text{if } X'_{n/2} \in \Re_0 \\ \psi_1(X'_{n-k'}) \cdot X''_{k'} & \text{if } X'_{n-k'} \in \Re_1. \end{cases} \quad (37)$$

**Theorem 2:** Let $f' \in \mathbb{F}_2^n$ be as in Construction 2. Then $f$ satisfies SAC, and has the nonlinearity

$$N_{f'} = 2^{n-1} - 2^{n/2-1} - 2^{k'-1}. \quad (38)$$

Let $\psi_1(X'_{n-k'})_j$ be the $j$-th component of $\psi_1(X'_{n-k'})$, where $1 \leq j \leq k'$. If

$$\bigoplus_{X'_{n-k'} \in \Re_1} \phi_1(X'_{n-k'})_j = 1 \quad (39)$$

for some $j$, then the algebraic degree of $f'$ is $d = n - k' + 1$.

**Proof:** Similarly to the proof of Theorem 1, $f'$ is an almost optimal $(n, m, n - k' + 1, N_{f'})$ function with $N_{f'} = 2^{n-1} - 2^{n/2-1} - 2^{k'-1}$. Next we prove that $f'$ satisfies SAC.

$$\Delta_{f'}(\alpha) = \sum_{X_n \in \mathbb{F}_2^n} (-1)^{f(X_n) \oplus f(X_n + \alpha)} = U_0 + U_1 \quad (40)$$

where

$$U_0 = \sum_{X'_{n/2} \in \Re_0} \sum_{X''_{n/2} \in \mathbb{F}_2^{n/2}} (-1)^{\psi_0(X'_{n/2}) \cdot X''_{n/2} \oplus \psi_0(X'_{n/2} + \alpha'_{n/2}) \cdot (X''_{n/2} + \alpha''_{n/2})}$$

$$= \sum_{X'_{n/2} \in \Re_0} (-1)^{\psi_0(X'_{n/2} + \alpha'_{n/2}) \cdot \alpha''_{n/2}} \sum_{X''_{n/2} \in \mathbb{F}_2^{n/2}} (-1)^{(\psi_0(X'_{n/2}) + \psi_0(X'_{n/2} + \alpha'_{n/2})) \cdot X''_{n/2}} \quad (41)$$

$$U_1 = \sum_{X'_{n-k'} \in \Re_1} \sum_{X''_{k'} \in \mathbb{F}_2^{k'}} (-1)^{\psi_1(X'_{n-k'}) \cdot X''_{k'} \oplus \psi_1(X'_{n-k'} + \alpha'_{n-k'}) \cdot (X''_{k'} + \alpha''_{k'})}$$

$$= \sum_{X'_{n/2} \in \Re_1} (-1)^{\psi_1(X'_{n-k'} + \alpha'_{n-k'}) \cdot \alpha''_{k'}} \sum_{X''_{k'} \in \mathbb{F}_2^{k'}} (-1)^{(\psi_1(X'_{n-k'}) + \psi_1(X'_{n-k'} + \alpha'_{n-k'})) \cdot X''_{k'}} \quad (42)$$

When $wt(\alpha) = 1$, to compute $U_0$, there exists two cases to be considered:

Case 1: $wt(\alpha'_{n/2}) = 1$ and $wt(\alpha''_{n/2}) = 0$. Since $\alpha'_{n/2} \neq 0$ and $\psi_0$ is an bijection from $\Re_0$ to $\Gamma_0$, we have

$$\psi_0(X'_{n/2}) + \psi_0(X'_{n/2} + \alpha'_{n/2}) \neq 0 \quad (43)$$

It follows that

$$\sum_{X''_{n/2} \in \mathbb{F}_2^{n/2}} (-1)^{(\psi_0(X'_{n/2}) + \psi_0(X'_{n/2} + \alpha'_{n/2})) \cdot X''_{n/2}} = 0 \quad (44)$$

Then, $U_0 = 0$.

Case 2: $wt(\alpha'_{n/2}) = 0$ and $wt(\alpha''_{n/2}) = 1$. In this case,

$$U_0 = 2^{n/2} \cdot \sum_{X'_{n/2} \in \Re_0} (-1)^{\psi_0(X'_{n/2}) \cdot \alpha''_{n/2}} \tag{45}$$

Due to the fact that for any $\beta \in \Omega$, $\beta^c \in \Omega$, we have

$$|\{X'_{n/2} \mid \psi_0(X'_{n/2}) \cdot \alpha''_{n/2} = 0, X'_{n/2} \in \Re_0\}| = |\{X'_{n/2} \mid \psi_0(X'_{n/2}) \cdot \alpha''_{n/2} = 1, X'_{n/2} \in \Re_0\}| \tag{46}$$

Thus, $U_0 = 0$.

So, $U_0 = 0$ when $wt(\alpha) = 1$. Similarly, $U_1 = 0$ when $wt(\alpha) = 1$. Hence, $\Delta_{f'}(\alpha) = 0$ when $wt(\alpha) = 1$. $f'$ satisfies SAC. $\square$

### 3.4 Degree optimization

The algebraic degree of any $(n, m, d, N_f)$ function $f$ obtained in Construction 1 can be optimized by adding a monomial $x_{i_{m+2}} \cdots x_{i_{k''}}$ to one subfunction

$$g = \phi_1(\delta) \cdot X''_{k''} = x_{i_1} \oplus x_{i_2} \oplus \cdots \oplus x_{i_{m+1}} \oplus l(x_{i_{m+2}}, x_{i_{m+3}}, \cdots, x_{i_{k''}}) \tag{47}$$

where $\delta \in E_1$ and $l \in \mathcal{B}_{k''-m-1}$ is a linear function. It is not difficult to prove that the nonlinearity of the degree optimized function $f''$ is equal to $N_f$, or $N_f - 2^{m+1}$. To ensure that $N_{f''} = N_f$ under certain condition, we below propose a method to optimize the algebraic degree of the GMM functions. This idea has been considered by Pasalic in [17], and later also be used in [1].

**Construction 3:** Let $n \geq 12$ be an even number, $m$ be a positive integer such that there exists an integer $k''$ with

$$k'' = \min_{m < s < n/2}\{s \mid 2^{n/2-s} \cdot \sum_{i=0}^{m} \binom{n/2}{i} \leq \sum_{j=m+1}^{s} \binom{s}{j} - 2^{s-m-1} + 1\}. \tag{48}$$

Let

$$\{i_1, i_2, \cdots, i_{m+1}\} \cup \{i_{m+2}, i_{m+3}, \cdots, i_{k''}\} = \{n - k'' + 1, n - k'' + 2, \cdots, n\}$$

and

$$T'_1 = \{c \mid c \in \mathbb{F}_2^{k''}, wt(c) > m, (c_{i_1}, c_{i_2}, \cdots, c_{i_{m+1}}) \neq (11 \cdots 1)\} \tag{49}$$

where $c = (c_{n-k''+1}, c_{n-k''+2}, \cdots, c_n)$. $E_0$, $T_1$, and $\phi_0$ are defined as in Construction 1. Let $E'_1 = \overline{E_0} \times \mathbb{F}_2^{n/2-k''}$. For any fixed $\delta \in E'_1$, $\phi'_1$ is any injective mapping from $E'_1 \setminus \{\delta\}$ to $T'_1$, and $\phi''_1$ any mapping from $\{\delta\}$ to $T_1 \setminus T'_1$. We construct the function $f'' \in \mathcal{B}_n$ as follows:

$$f''(X_n) = \begin{cases} \phi_0(X'_{n/2}) \cdot X''_{n/2}, & X'_{n/2} \in E_0 \\ \phi'_1(X'_{n-k''}) \cdot X''_{k''}, & X'_{n-k''} \in E'_1 \setminus \{\delta\} \\ \phi''_1(X'_{n-k''}) \cdot X''_{k''} + x_{i_{m+2}} \cdots x_{i_{k''}}, & X'_{n-k''} = \delta. \end{cases} \tag{50}$$

**Theorem 3:** A function $f'' \in \mathbb{F}_2^n$ is proposed by Construction 3. Then $f''$ is an almost optimal $(n, m, n - m - 1, N_{f''})$ function with

$$N_{f''} = 2^{n-1} - 2^{n/2-1} - 2^{k''-1}. \tag{51}$$

**Proof:** Since the term $x_1 \cdots x_{n-k''} x_{i_{m+2}} \cdots x_{i_{k''}}$ appears in the ANF of $f''$, we have

$$\deg(f'') = n - m - 1.$$

For any $\omega = (\omega_1, \cdots, \omega_n) \in \mathbb{F}_2^n$ we have $W_{f''}(\omega) = S_0 + S_1' + S_1''$ where

$$S_0 = \sum_{X'_{n/2} \in E_0} (-1)^{(\omega_1,\cdots,\omega_{n/2}) \cdot X'_{n/2}} \sum_{X''_{n/2} \in \mathbb{F}_2^{n/2}} (-1)^{(\phi_0(X'_{n/2}) + (\omega_{n/2+1},\cdots,\omega_n)) \cdot X''_{n/2}}$$

$$= \begin{cases} 0, & 0 \leq wt(\omega_{n/2+1}, \cdots, \omega_n) \leq m \\ \pm 2^{n/2}, & m+1 \leq wt(\omega_{n/2+1}, \cdots, \omega_n) \leq n/2 \end{cases} \quad (52)$$

$$S_1' = \sum_{X'_{n-k} \in E_1' \setminus \{\delta\}} (-1)^{(\omega_1,\cdots,\omega_{n-k''}) \cdot X'_{n-k''}} \sum_{X''_{k''} \in \mathbb{F}_2^{k''}} (-1)^{(\phi_1(X'_{n-k''}) + (\omega_{n-k''+1},\cdots,\omega_n)) \cdot X''_{k''}}$$

$$= \begin{cases} 0, & 0 \leq wt(\omega_{n-k''+1}, \cdots, \omega_n) \leq m \\ \pm 2^{k''}, & m+1 \leq wt(\omega_{n-k''+1}, \cdots, \omega_n) \leq n/2 \end{cases} \quad (53)$$

$$S_1'' = (-1)^{(\omega_1,\cdots,\omega_{n-k''}) \cdot \delta} \sum_{X''_{k''} \in \mathbb{F}_2^{k''}} (-1)^{(\phi_1''(\delta) + (\omega_{n-k''+1},\cdots,\omega_n)) \cdot X''_{k''} + x_{i_{m+2}} \cdots x_{i_{k''}}}$$

$$= \begin{cases} 0, & (\omega_{n-k''+1}, \cdots, \omega_n) \neq \phi_1''(\delta), (\omega_{i_1}, \cdots, \omega_{i_{m+1}}) \neq (1 \cdots, 1) \\ \pm 2^{m+2}, & (\omega_{n-k''+1}, \cdots, \omega_n) \neq \phi_1''(\delta), (\omega_{i_1}, \cdots, \omega_{i_{m+1}}) = (1 \cdots, 1) \\ \pm(2^{k''} - 2^{m+2}), & (\omega_{n-k''+1}, \cdots, \omega_n) = \phi_1''(\delta). \end{cases} \quad (54)$$

Then clearly,

$$S_1' + S_1'' = \begin{cases} 0, & 0 \leq wt(\omega_{n-k''+1}, \cdots, \omega_n) \leq m \\ \pm 2^{m+2}, & (\omega_{n-k''+1}, \cdots, \omega_n) \neq \phi_1''(\delta), (\omega_{i_1}, \cdots, \omega_{i_{m+1}}) = (1 \cdots, 1) \\ \pm(2^{k''} - 2^{m+2}), & (\omega_{n-k''+1}, \cdots, \omega_n) = \phi_1''(\delta) \\ \pm 2^{k''}, & \text{else.} \end{cases} \quad (55)$$

So we have

$$\max_{\omega \in \mathbb{F}_2^n} |W_{f''}(\omega)| = 2^{n/2} + 2^{k''}.$$

From (4),

$$N_{f''} = 2^{n-1} - 2^{n/2-1} - 2^{k''-1}.$$

When $0 \leq wt(\omega) \leq m$, we always have

$$0 \leq wt(\omega_{n/2+1}, \cdots, \omega_n) \leq m$$

and

$$0 \leq wt(\omega_{n-k''+1}, \cdots, \omega_n) \leq m.$$

From (52) and (55), $W_{f''}(\omega) = 0$. By Lemma 1, $f''$ is an $m$-resilient function. $\square$

## 4 Construction of almost optimal $m$-resilient functions on $n$ variables ($n$ odd) with nonlinearity $> 2^{n-1} - 2^{(n-1)/2}$

For odd $n$, 15-variable Boolean functions with nonlinearity 16276 were constructed by Patterson and Wiedemann (PW) [22]. Recently, 9-variable Boolean functions with nonlinearity 242 were found by Kavut and Yücel (KY) [21]. We will use PW functions (or KY functions) to construct $m$-resilient functions with nonlinearity greater than $> 2^{n-1} - 2^{(n-1)/2}$ for odd $n$.

**Theorem 4:** Let $n = n_0 + 15$ (respectively $n = n_0 + 9$) where $n_0$ be even, and $m, k$ be positive integers such that

$$k = \min_{m < s < n_0/2} \{s \mid 2^{n_0/2 - s} \cdot \sum_{i=0}^{m} \binom{n_0/2}{i} \leq \sum_{j=m+1}^{s} \binom{s}{j}\} \leq n_0/2 - 3. \tag{56}$$

It is possible to construct an almost optimal $m$-resilient functions $f \in \mathcal{B}_n$ with

$$N_f = 2^{n-1} - 2^{(n-1)/2} + 5 \cdot 2^{n_0/2+2} - 27 \cdot 2^{k+2}$$

(respectively $\quad N_f = 2^{n-1} - 2^{(n-1)/2} + 2^{n_0/2+1} - 7 \cdot 2^{k+1}$)

**Proof:** If (56) holds, then we can construct an $(n_0, m, -, 2^{n_0-1} - 2^{n_0/2-1} - 2^{k-1})$ function $f_0 \in \mathcal{B}_{n_0}$ by the method proposed in Construction 1 (Note that examples can be found in Appendix 1). Let $g \in \mathcal{B}_{15}$ be a PW function, and $f \in \mathcal{B}_n$ defined by

$$f(X_n) = f_0(X'_{n_0}) \oplus g(X''_{15}).$$

We can easily deduce that

$$\begin{aligned} N_f &= 2^{n-1} - 1/2 \cdot (2^{n_0} - 2N_{f_0})(2^{15} - 2N_g) \\ &= 2^{n-1} - 2^{(n-1)/2} + 5 \cdot 2^{n_0/2+2} - 27 \cdot 2^{k+2} \\ &> 2^{n-1} - 2^{(n-1)/2}. \end{aligned}$$

When $g \in \mathcal{B}_9$ be a KY function, the proof is similar. □

Let $n_m$ be the minimum $n_0$ such that the nonlinearity of the $m$-resilient functions $f \in \mathcal{B}_{n_m+15}$ (or $f \in \mathcal{B}_{n_m+9}$) constructed above is strictly greater than $2^{n-1} - 2^{(n-1)/2}$. By using the information in Appendix 1, we have

| $m$ | 1 | 2 | 3 | 4 | 5 | 6 | 7 | 8 | 9 | 10 |
|---|---|---|---|---|---|---|---|---|---|---|
| $n_m$ | 20 | 26 | 32 | 38 | 42 | 48 | 52 | 58 | 62 | 68 |
| $n_m + 15$ | 35 | 41 | 47 | 53 | 57 | 63 | 67 | 73 | 77 | 83 |
| $n_m + 9$ | 29 | 35 | 41 | 47 | 51 | 57 | 61 | 67 | 71 | 77 |

## 5 Construction of multiple-output almost optimal resilient functions $F : \mathbb{F}_2^n \mapsto \mathbb{F}_2^r$ ($n$ even) with nonlinearity $> 2^{n-1} - 2^{n/2}$

Constructing multiple-output resilient functions with high nonlinearity has received attention since mid-1990s [23][24][25][26][27][28][29][30]. To the best of our knowledge, the nonlinearity of the multiple-output resilient functions on $\mathbb{F}_2^n$ obtained by the existing constructions is at most $2^{n-1} - 2^{\lfloor n/2 \rfloor}$. In this section, we present a technique on constructing an $m$-resilient function, $F : \mathbb{F}_2^n \mapsto \mathbb{F}_2^r$ ($n$ even) with nonlinearity $2^{n-1} - 2^{n/2-1} - 2^{k-1}$ where $k < n/2$.

**Definition 3:** The nonlinearity of $F = (f_1, f_2, \cdots, f_r)$, denoted by $N_F$, is defined as [31]

$$N_F = \min_{c \in \mathbb{F}_2^r \setminus \{0\}} N_{f_c}$$

where $f_c = \sum_{i=1}^r c_i f_i$, $c = (c_1, \cdots, c_r) \in \mathbb{F}_2^r$. $F$ is said to be almost optimal if $N_F \geq 2^{n-1} - 2^{\lfloor n/2 \rfloor}$.

**Lemma 2 ([24]):** A function $F = (f_1, f_2, \cdots, f_r)$ is an $m$-resilient function if and only if for any $c = (c_1, \cdots, c_r) \in \mathbb{F}_2^r \setminus \{0\}$, $f_c = \sum_{i=1}^r c_i f_i$ is an $m$-resilient function.

**Definition 4 ([27]):** A set of $[n, k]$ linear codes $\{C_1, C_2, \cdots, C_s\}$ such that

$$C_i \cap C_j = \{0\}, \quad 1 \leq i < j \leq s \tag{57}$$

is called a set of $[n, k]$ disjoint linear codes. Let $d_i$ be the minimum weight of the nonzero code vectors in $C_i$, $0 \leq i \leq s$. $\{C_1, C_2, \cdots, C_s\}$ is called a set of $[n, k, \geq d^*]$ disjoint linear codes, where $d^* = \min\{d_1, d_2, \cdots, d_s\}$.

**Construction 4:** Let $n \geq 12$ be even and $r, m \leq \lfloor n/4 \rfloor$ be positive integers. Let $C = \{C_1, \cdots, C_u\}$ be a set of $[n/2, r, \geq m+1]$ disjoint linear codes with $u$ as large as possible, and associate to each code a mapping $\rho_i : \mathbb{F}_{2^r} \mapsto C_i, 1 \leq i \leq u$, so that

$$(b_0, b_1 \alpha, \cdots b_{m-1} \alpha^{m-1}) \xrightarrow{\rho_i} b_0 \theta_0^i + \cdots + b_{m-1} \theta_{m-1}^i \tag{58}$$

where $\alpha$ is primitive in $\mathbb{F}_{2^r}$ and $\theta_0^i, \cdots, \theta_{m-1}^i$ is a basis of $C_i$. Define the matrix $A_i$ by

$$A_i = \begin{pmatrix} \rho_i(1) & \rho_i(\alpha) & \cdots & \rho_i(\alpha^{r-1}) \\ \rho_i(\alpha) & \rho_i(\alpha^2) & \cdots & \rho_i(\alpha^r) \\ \vdots & \vdots & \ddots & \vdots \\ \rho_i(\alpha^{2^r-2}) & \rho_i(1) & \cdots & \rho_i(\alpha^{r-2}) \end{pmatrix}$$

Let $C' = \{C'_1, \cdots, C'_v\}$ be a set of $[s, r, \geq m+1]$ disjoint linear codes with $v$ as large as possible, and associate to each code a mapping $\varrho_j : \mathbb{F}_{2^r} \mapsto C'_j, 1 \leq j \leq v$, so that

$$(b_0, b_1 \beta, \cdots b_{m-1} \beta^{m-1}) \xrightarrow{\varrho_j} b_0 \eta_0^j + \cdots + b_{m-1} \eta_{m-1}^j \tag{59}$$

where $\beta$ is primitive in $\mathbb{F}_{2^r}$ and $\eta_0^j, \cdots, \eta_{m-1}^j$ is a basis of $C'_j$. Define the matrix $B_j$ by

$$B_j = \begin{pmatrix} \varrho_j(1) & \varrho_j(\beta) & \cdots & \varrho_j(\beta^{r-1}) \\ \varrho_j(\beta) & \varrho_j(\beta^2) & \cdots & \varrho_j(\beta^r) \\ \vdots & \vdots & \ddots & \vdots \\ \varrho_j(\beta^{2^r-2}) & \varrho_j(1) & \cdots & \varrho_j(\beta^{r-2}) \end{pmatrix}$$

We define

$$k = \min_{m < s < n/2} \{s \mid 2^{n/2-s} \cdot (2^{n/2} - u \cdot (2^r - 1)) \leq v \cdot (2^r - 1)\} \tag{60}$$

Let $E_0 = \{e_1, e_2, \cdots, e_\kappa\}$ be any subset of $\mathbb{F}_2^{n/2}$ with $\kappa = |E_0| = u \cdot (2^r - 1)$. Let $\overline{E_0} = \mathbb{F}_2^{n/2} \setminus E_0$ and $E_1 = \overline{E_0} \times \mathbb{F}_2^{n/2-k} = \{\epsilon_1, \epsilon_2, \cdots, \epsilon_\lambda\}$ with $\lambda = 2^{n/2-s} \cdot (2^{n/2} - u \cdot (2^r - 1))$. Define $T_0 = A_1 \cup A_2 \cup \cdots \cup A_u$ and $T_1 = B_1 \cup B_2 \cup \cdots \cup B_u$. For $1 \leq i \leq r$, let $\psi_i$ be an injective mapping from $E_0$ to $T_0$ such that

$$\begin{pmatrix} \psi_1(e_1) & \psi_2(e_1) & \ldots & \psi_r(e_1) \\ \psi_1(e_2) & \psi_2(e_2) & \ldots & \psi_r(e_2) \\ \vdots & \vdots & \ddots & \vdots \\ \psi_1(e_\kappa) & \psi_2(e_\kappa) & \ldots & \psi_r(e_\kappa) \end{pmatrix} = (A_1^T | A_2^T | \cdots | A_u^T)^T.$$

Let $\varphi_i$ be an injective mapping from $E_1$ to $T_1$ such that

$$\begin{pmatrix} \varphi_1(\epsilon_1) & \varphi_2(\epsilon_1) & \ldots & \varphi_r(\epsilon_1) \\ \varphi_1(\epsilon_2) & \varphi_2(\epsilon_2) & \ldots & \varphi_r(\epsilon_2) \\ \vdots & \vdots & \ddots & \vdots \\ \varphi_1(\epsilon_\lambda) & \varphi_2(\epsilon_\lambda) & \ldots & \varphi_r(\epsilon_\lambda) \end{pmatrix} = \overline{(B_1^T | B_2^T | \cdots | B_u^T)^T}$$

where $\overline{(B_1^T | B_2^T | \cdots | B_u^T)^T}$ denotes that some rows of $(B_1^T | B_2^T | \cdots | B_u^T)^T$ may be deleted to be of size $\lambda \times r$. We now construct the a function $F : \mathbb{F}_2^n \mapsto \mathbb{F}_2^r$ by

$$F(X_n) = (f_1(X_n), f_2(X_n), \cdots, f_r(X_n))$$

where

$$f_i(X_n) = \begin{cases} \psi_i(X'_{n/2}) \cdot X''_{n/2} & \text{if } X'_{n/2} \in E_0 \\ \varphi_i(X'_{n-k}) \cdot X''_k & \text{if } X'_{n-k} \in E_1 \end{cases} \quad i = 1, 2, \cdots r. \tag{61}$$

**Theorem 5:** Let $F : \mathbb{F}_2^n \mapsto \mathbb{F}_2^r$ be as in Construction 4. Then $F$ is an almost optimal $m$-resilient function with

$$N_F = 2^{n-1} - 2^{n/2-1} - 2^{k-1}. \tag{62}$$

**Proof:** Let $\psi_c = c_1\psi_1 + c_2\psi_2 + \cdots + c_r\psi_r$ where $c = (c_1, \cdots, c_r) \in \mathbb{F}_2^r \setminus \{0\}$. For $i = 1, 2, \cdots, r$, $\psi_i$ is injective, it is not difficult to prove that $\psi_c$ is injective. Similarly, $\varphi_c = c_1\varphi_1 + c_2\varphi_2 + \cdots + c_r\varphi_r$ is injective. Let $\alpha = (\beta', \beta'') = (\gamma', \gamma'') \in \mathbb{F}_2^n$, where $\gamma' \in \mathbb{F}_2^{n-k}$, $\gamma'' \in \mathbb{F}_2^k$, and $\beta', \beta'' \in \mathbb{F}_2^{n/2}$. Then

$$W_{f_c}(\alpha) = \sum_{X_n \in \mathbb{F}_2^n} (-1)^{f_c(X_n) \oplus \alpha \cdot X_n} = U_0 + U_1 \tag{63}$$

where

$$U_0 = \sum_{X'_{n/2} \in E_0} \sum_{X''_{n/2} \in \mathbb{F}_2^{n/2}} (-1)^{\psi_c(X'_{n/2}) \cdot X''_{n/2} \oplus (\beta', \beta'') \cdot (X'_{n/2}, X''_{n/2})} \tag{64}$$

$$= \sum_{X'_{n/2} \in E_0} (-1)^{\beta' \cdot X'_{n/2}} \sum_{X''_{n/2} \in \mathbb{F}_2^{n/2}} (-1)^{(\psi_c(X'_{n/2}) + \beta'') \cdot X''_{n/2}} \tag{65}$$

and

$$U_1 = \sum_{X'_{n-k} \in E_1} (-1)^{\gamma' \cdot X'_{n-k}} \sum_{X''_k \in \mathbb{F}_2^k} (-1)^{(\varphi_c(X'_{n-k}) + \gamma'') \cdot X''_k} \tag{66}$$

When $\psi^{-1}(\beta'') = \emptyset$, we have $U_0 = 0$; or else

$$U_0 = 2^{n/2} \cdot (-1)^{\beta'} \cdot \psi^{-1}(\beta'') = \pm 2^{n/2}.$$

Similarly,
$$U_1 \in \{0, \pm 2^k\}.$$

We have
$$W_{f_c} \in \{0, \pm 2^k, \pm 2^{n/2}, \pm(2^{n/2} - 2^k), \pm(2^{n/2} + 2^k)\}.$$

Then
$$N_{f_c} = 2^{n-1} - 2^{n/2-1} - 2^{k-1}.$$

From Definition 2,
$$N_F = 2^{n-1} - 2^{n/2-1} - 2^{k-1}.$$

Noticing that for any $\beta'' \in T_0$, $\gamma'' \in T_1$, we always have $wt(\beta'') \geq m+1$ and $wt(\gamma'') \geq m+1$. With the similar proof as in Theorem 1 (see Case 1), we can obtain that $f_c$ is an $m$-resilient function. By Lemma 2, $F$ is an $m$-resilient function. $\square$

## 6 Conclusion and open problems

In this paper, we present a generalized Maiorana-McFarland (GMM) construction method to obtain almost optimal resilient functions with a nonlinearity higher than that attainable by any previously known construction method. The following problems are left for future work.

*Conjectures:*

1) Let $n \geq 12$ be even and $m < \lceil n/4 \rceil$. For any $(n, m, -, N_f)$ function $f \in \mathcal{B}_n$, $N_f \leq 2^{n-1} - 2^{n/2-1} - 2^{\lfloor n/4 \rfloor + m - 1}$.

2) Let $n \geq 12$ be even and $\lceil n/4 \rceil \leq m \leq n/2 - 2$. For any $(n, m, -, N_f)$ function $f \in \mathcal{B}_n$, $N_f < 2^{n-1} - 2^{n/2-1} - 2^{m+1}$.

3) Let $n \geq 12$ be even and $m \leq n/2 - 2$. If $F : \mathbb{F}_2^n \mapsto \mathbb{F}_2^r$ is a multiple-output function with nonlinearity $N_F > 2^{n-1} - 2^{n/2}$, then $m + r \leq n/2 - 1$.

**Appendix 1:** Examples of $(n, m, n-k+1, 2^{n-1} - 2^{n/2-1} - 2^{k-1})$ resilient functions.

| | | | | | | | | | | | | | | | | | | |
|---|---|---|---|---|---|---|---|---|---|---|---|---|---|---|---|---|---|---|
| | | | | | | | | | $m=1$ | | | | | | | | | |
| $n$ | 12 | 16 | 20 | 24 | 28 | 32 | 34 | 36 | 38 | 42 | 46 | 50 | 54 | 58 | 62 | 64 | 66 | 68 |
| $k$ | 5 | 6 | 7 | 8 | 9 | 11 | 11 | 12 | 12 | 13 | 14 | 15 | 16 | 17 | 19 | 19 | 20 | 20 |
| $n$ | 70 | 72 | 74 | 76 | 80 | 84 | 88 | 92 | 96 | 100 | 128 | 250 | 500 | 600 | 1000 | 5000 | 10000 | 40000 |
| $k$ | 21 | 21 | 22 | 22 | 23 | 24 | 25 | 26 | 27 | 28 | 36 | 66 | 129 | 155 | 255 | 1256 | 2507 | 10008 |
| | | | | | | | | | $m=2$ | | | | | | | | | |
| $n$ | 16 | 18 | 20 | 22 | 26 | 30 | 32 | 34 | 36 | 40 | 44 | 46 | 48 | 50 | 54 | 58 | 62 | 64 |
| $k$ | 7 | 8 | 9 | 9 | 10 | 11 | 12 | 13 | 13 | 14 | 15 | 16 | 17 | 17 | 18 | 19 | 20 | 21 |
| $n$ | 66 | 68 | 70 | 72 | 76 | 80 | 84 | 88 | 90 | 92 | 94 | 96 | 98 | 100 | 250 | 500 | 1000 | 10000 |
| $k$ | 22 | 22 | 23 | 23 | 24 | 25 | 26 | 27 | 28 | 29 | 29 | 30 | 30 | 31 | 66 | 133 | 259 | 2512 |
| | | | | | | | | | $m=3$ | | | | | | | | | |
| $n$ | 20 | 22 | 26 | 28 | 30 | 32 | 36 | 38 | 42 | 44 | 46 | 48 | 52 | 56 | 58 | 60 | 62 | 66 |
| $k$ | 9 | 10 | 11 | 12 | 13 | 13 | 14 | 15 | 16 | 17 | 18 | 18 | 19 | 20 | 21 | 22 | 22 | 23 |
| $n$ | 70 | 72 | 74 | 76 | 78 | 80 | 84 | 88 | 92 | 94 | 96 | 98 | 100 | 198 | 250 | 500 | 1000 | 10000 |
| $k$ | 24 | 25 | 26 | 26 | 27 | 27 | 28 | 29 | 30 | 31 | 32 | 32 | 33 | 59 | 72 | 136 | 263 | 2518 |
| | | | | | | | | | $m=4$ | | | | | | | | | |
| $n$ | 26 | 28 | 30 | 32 | 34 | 36 | 38 | 42 | 44 | 48 | 50 | 52 | 54 | 58 | 60 | 62 | 64 | 68 |
| $k$ | 12 | 13 | 14 | 14 | 15 | 16 | 16 | 17 | 18 | 19 | 20 | 21 | 21 | 22 | 23 | 24 | 24 | 25 |
| $n$ | 70 | 72 | 74 | 76 | 82 | 84 | 86 | 88 | 92 | 96 | 100 | 122 | 148 | 200 | 250 | 500 | 1000 | 10000 |
| $k$ | 26 | 27 | 27 | 28 | 29 | 30 | 31 | 31 | 32 | 33 | 34 | 41 | 48 | 61 | 75 | 139 | 266 | 2523 |
| | | | | | | | | | $m=5$ | | | | | | | | | |
| $n$ | 30 | 32 | 34 | 36 | 38 | 40 | 42 | 44 | 46 | 48 | 50 | 52 | 54 | 56 | 58 | 60 | 62 | 64 |
| $k$ | 14 | 15 | 16 | 16 | 17 | 18 | 18 | 19 | 20 | 20 | 21 | 22 | 22 | 23 | 24 | 24 | 25 | 25 |
| $n$ | 66 | 70 | 72 | 74 | 76 | 80 | 84 | 86 | 88 | 90 | 94 | 96 | 98 | 100 | 250 | 500 | 1000 | 10000 |
| $k$ | 26 | 27 | 28 | 28 | 29 | 30 | 31 | 32 | 33 | 33 | 34 | 35 | 36 | 36 | 77 | 142 | 269 | 2523 |
| | | | | | | | | | $m=6$ | | | | | | | | | |
| $n$ | 34 | 40 | 42 | 46 | 48 | 52 | 54 | 58 | 60 | 64 | 66 | 70 | 76 | 80 | 82 | 86 | 90 | 100 |
| $k$ | 16 | 19 | 19 | 21 | 21 | 23 | 23 | 25 | 25 | 27 | 27 | 28 | 30 | 32 | 32 | 33 | 35 | 38 |
| | | | | | | | | | $m=7$ | | | | | | | | | |
| $n$ | 38 | 44 | 46 | 52 | 58 | 60 | 64 | 66 | 70 | 72 | 76 | 82 | 86 | 88 | 92 | 98 | 100 | 102 |
| $k$ | 18 | 21 | 21 | 23 | 26 | 26 | 28 | 28 | 30 | 30 | 31 | 33 | 35 | 35 | 36 | 38 | 39 | 40 |
| | | | | | | | | | $m=8$ | | | | | | | | | |
| $n$ | 44 | 50 | 56 | 58 | 64 | 70 | 76 | 82 | 88 | 92 | 94 | 98 | 100 | 200 | 248 | 250 | 500 | 1000 |
| $k$ | 21 | 23 | 26 | 26 | 28 | 30 | 32 | 34 | 36 | 38 | 38 | 40 | 40 | 69 | 83 | 83 | 150 | 279 |
| | | | | | | | | | $m=9$ | | | | | | | | | |
| $n$ | 48 | 54 | 56 | 62 | 68 | 70 | 76 | 82 | 88 | 94 | 98 | 100 | 104 | 150 | 250 | 252 | 500 | 100 |
| $k$ | 23 | 26 | 26 | 28 | 31 | 31 | 33 | 35 | 37 | 39 | 41 | 41 | 43 | 57 | 85 | 86 | 152 | 282 |
| | | | | | | | | | $m=10$ | | | | | | | | | |
| $n$ | 52 | 60 | 66 | 68 | 74 | 80 | 82 | 86 | 88 | 94 | 100 | 148 | 152 | 200 | 250 | 300 | 500 | 1000 |
| $k$ | 25 | 28 | 31 | 31 | 33 | 36 | 36 | 38 | 38 | 40 | 42 | 57 | 59 | 73 | 87 | 101 | 154 | 284 |
| | | | | | | | | | $m=100$ | | | | | | | | | |
| $n$ | 428 | 500 | 600 | 1000 | 2000 | 3000 | 4000 | 5000 | 6000 | 7000 | 8000 | 10000 | 20000 | 30000 | 40000 | 50000 | | |
| $k$ | 213 | 245 | 287 | 429 | 733 | 1013 | 1285 | 1551 | 1814 | 2076 | 2336 | 2852 | 5402 | 7932 | 10452 | 12968 | | |

***Appendix 2:*** Examples of GMM resilient functions, where $n$ is the minimum value such that the nonlinearity of an $m$-resilient function is $2^{n-1} - 2^{n/2-1} - 2^{n/2-2}$.

| $N_f = 2^{n-1} - 2^{n/2-1} - 2^{n/2-2}$ | | | | | | | | | | | | | | |
|---|---|---|---|---|---|---|---|---|---|---|---|---|---|---|
| $m$ | 1 | 2 | 3 | 4 | 5 | 6 | 7 | 8 | 9 | 10 | 11 | 12 | 13 | 14 | 15 |
| $n$ | 12 | 16 | 20 | 12 | 30 | 34 | 38 | 44 | 48 | 52 | 56 | 60 | 64 | 70 | 74 |
| $m$ | 16 | 17 | 18 | 19 | 20 | 21 | 22 | 23 | 24 | 25 | 26 | 27 | 28 | 29 | 30 |
| $n$ | 78 | 82 | 86 | 90 | 94 | 100 | 104 | 108 | 112 | 116 | 120 | 122 | 128 | 134 | 138 |
| $m$ | 31 | 32 | 33 | 34 | 35 | 36 | 37 | 38 | 39 | 40 | 41 | 42 | 43 | 44 | 45 |
| $n$ | 142 | 144 | 150 | 154 | 158 | 162 | 166 | 170 | 176 | 180 | 184 | 188 | 192 | 196 | 200 |
| $m$ | 46 | 47 | 48 | 49 | 50 | 51 | 52 | 53 | 54 | 55 | 56 | 57 | 58 | 59 | 60 |
| $n$ | 204 | 208 | 212 | 216 | 222 | 226 | 230 | 234 | 238 | 242 | 246 | 250 | 254 | 258 | 262 |
| $m$ | 61 | 62 | 63 | 64 | 65 | 66 | 67 | 68 | 69 | 70 | 71 | 72 | 73 | 74 | 75 |
| $n$ | 266 | 270 | 276 | 280 | 284 | 288 | 292 | 296 | 300 | 304 | 308 | 312 | 316 | 320 | 324 |
| $m$ | 76 | 77 | 78 | 79 | 80 | 81 | 82 | 83 | 84 | 85 | 86 | 87 | 88 | 89 | 90 |
| $n$ | 326 | 334 | 338 | 342 | 346 | 350 | 354 | 358 | 362 | 366 | 370 | 374 | 378 | 382 | 386 |
| $m$ | 91 | 92 | 93 | 94 | 95 | 96 | 97 | 98 | 99 | 100 | | | | | |
| $n$ | 390 | 396 | 400 | 404 | 408 | 412 | 416 | 420 | 424 | 428 | | | | | |